# *IMPACT OF ION CLEARING ELECTRODES ON BEAM DYNAMICS IN DAΦNE

M. Zobov, A. Battisti, A. Clozza, V. Lollo, C. Milardi,
B. Spataro, A. Stella, C. Vaccarezza, LNF-INFN, Frascati, Italy

* Work partly supported by the EC under the FP6 Research Infrastructure Action - Structuring the European Research Area EUROTeV DS Project Contract no.011899 RIDS

*Abstract*

Presently clearing electrodes are being considered as a possible cure of e-cloud driven problems in existing and future colliders. "Invisible" electrodes, made of a thin highly resistive layer pasted on a dielectric plate, have been proposed as one of design solutions for the e-cloud clearing. For the first time such electrodes were successfully used in the electron-positron accumulator (EPA) of LEP. Similar electrodes had been using for a long time for ion clearing purposes in the DAΦNE electron ring. Theoretical considerations and experimental measurements at DAΦNE have revealed a substantial contribution of the ion clearing electrodes (ICE) to the machine broad-band impedance giving rise to several harmful effects degrading the collider performance. In this paper we discuss the impact of the electrodes on DAΦNE beam dynamics, show the results of ICE wake field and impedance calculations and compare them with available experimental data. We also describe the procedure of ICE removal from the wiggler sections of the electron ring that has resulted in remarkable improvements in terms of beam dynamics and geometric luminosity.

## 1. INTRODUCTION

The Φ-factory DAΦNE [1] is an electron-positron collider at the energy of Φ resonance (1.02 GeV in the center of mass) designed and built in Frascati National Laboratories of INFN (see Fig. 1).

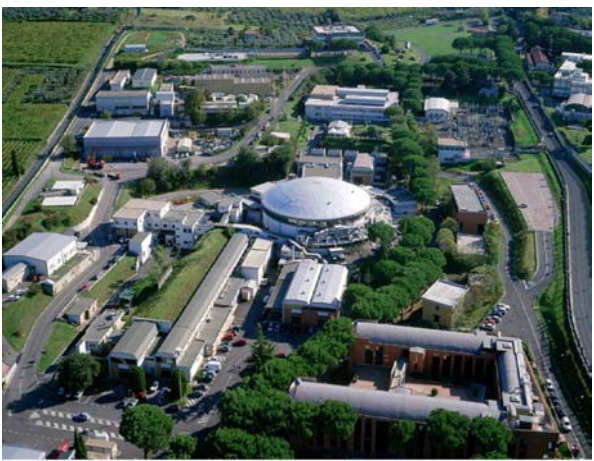

Fig. 1 View of DAΦNE accelerator complex.

The DAΦNE complex consists of two independent rings having two common interaction regions and an injection system composed of a full energy linear accelerator, a damping/accumulator ring and the relative transfer lines. The maximum peak luminosity obtained so far at DAΦNE is $1.6 \times 10^{32}$ cm$^{-2}$s$^{-1}$, while the maximum daily integrated luminosity is 10 pb$^{-1}$ [2]. Some of the main collider parameters are listed in Table 1.

In order to prevent accumulation of trapped ions in the electron beam potential several techniques were investigated. It was concluded that the DAΦNE start-up configuration requires a system of ion clearing electrodes (ICE) [3]. Because of the high circulating current stored in the electron multibunch beam (see Table 1) the Machine Advisory Committee suggested to use "invisible" electrodes similar to that proposed by F. Caspers [4] and already successfully used in the electron-positron accumulator (EPA) of CERN [5]. The principal goal of using such electrodes was to make them transparent for the beam as much as possible in order to:

- avoid the ICE to act as an antenna. Otherwise, the ICE would intercept a large fraction of the beam image current with a possible excessive power heating of the ICE themselves and an eventual damage of external feedthroughs and electronics;
- eliminate the resonant part of the beam coupling impedance arising from a mismatch between the ICE structures with the external feedthroughs and loadings. This is particularly important for the high current multibunch DAΦNE operations without coupled bunch instabilities.

Table 1: DAΦNE main parameters (KLOE run)

| Energy [GeV] | 0.51 |
|---|---|
| Trajectory length [m] | 97.69 |
| RF frequency [MHz] | 368.26 |
| Harmonic number | 120 |
| Damping time, $\tau_E/\tau_x$ [ms] | 17.8/36.0 |
| Bunch length [cm] | 1-3 |
| Number of colliding bunches | 111 |
| Beta functions $\beta_x/\beta_y$ [m] | 1.6/0.017 |
| Emittance, $\varepsilon_x$ [mm·mrad] (KLOE) | 0.34 |
| Coupling [%] | 0.2-0.3 |
| Max. tune shifts | 0.03/0.04 |
| Max. beam current e-/e+ [A] | 2.4/1.4 |

However, during routine operations several effects harmful for the collider performance have been observed in beam dynamics of the electron beam. Similar effects have not been revealed or were less pronounced in the positron ring. Among the most offending effects are the following: strong lengthening of the electron bunches; lower microwave instability threshold; vertical beam size blow up depending on the bunch current and the RF voltage; quadrupole longitudinal bunch oscillations.

Since the vacuum chambers of the both rings are similar with the exception of the ICE in the electron ring, we have performed a more detailed study of the ICE coupling impedance including both numerical simulations and a comparison of the simulation results with available experimental data on bunch lengthening. These studies have led to the decision to remove the 2 m long ICE from the wiggler section vacuum chambers.

In this paper we describe our observations and experience in running DAΦNE with the ICE. In Section 2 we describe the DAΦNE ICE design while the principal harmful effects are summarized in Section 3. The results of the ICE impedance calculations and their experimental verification are discussed in Sections 4 and 5. Finally, Section 6 describes the wiggler section ICE removal procedure and shows some of the resulting improvements in beam dynamics and collider performance.

## 2. ICE DESIGN

For the "invisible" ICE a material with a high "resistivity per square" $R_0$ is required:

$$R_0 = \rho \frac{l}{S} = \rho \frac{l}{dl} = \frac{\rho}{d} \qquad (1)$$

As it is seen from (1), $R_0$ does not depend on the material length $l$ and it should have a high resistivity $\rho$ and a small thickness $d$. In this case we can expect that the skin depth $\delta_s$ will be much larger than the ICE thickness to provide the ICE transparency at RF frequencies.

The DAΦNE ICE are made of a highly resistive paste with a thickness $d$ of 25 μm and the resistivity per square $\rho/d$ of the order of $10^5$ Ω painted on a dielectric material with ε = 9 (alumina). The skin depth estimated at a typical bunch spectrum frequency of 1 GHz is:

$$\delta_s = \sqrt{\frac{2\rho c}{Z_0 \omega}} = 0.025 m \rightarrow 1000 d, \qquad (2)$$

where $c$ is the velocity of light, $Z_0$ the free space impedance and $\omega$ the angular frequency, i.e. the skin depth is much larger than the layer thickness thus making the layer practically transparent for the RF frequencies.

The width of all ICE is about 5 cm, but their length and assembly designs depend on their locations along the ring. The electrodes in the straight sections having a round cross-section with a diameter of 88 mm are typically 10-12 cm long (see Fig. 2). The arc vacuum chambers are more complicated at it is seen in Fig.3. In the arcs there are short ICE installed in the bending magnet sections with the octagonal cross-section and the longest ICE placed in the very flat wiggler vacuum chambers (120 x 20 mm$^2$). The wiggler section ICE are 2.14 m long and stay very close to the beam trajectory, at about 1 cm distance.

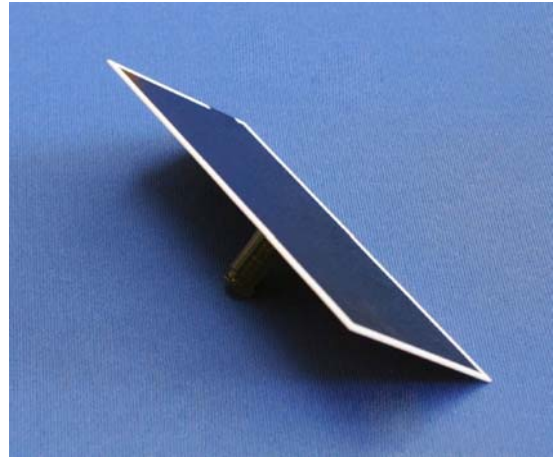

Fig. 2 Short ion clearing electrode located in straight sections.

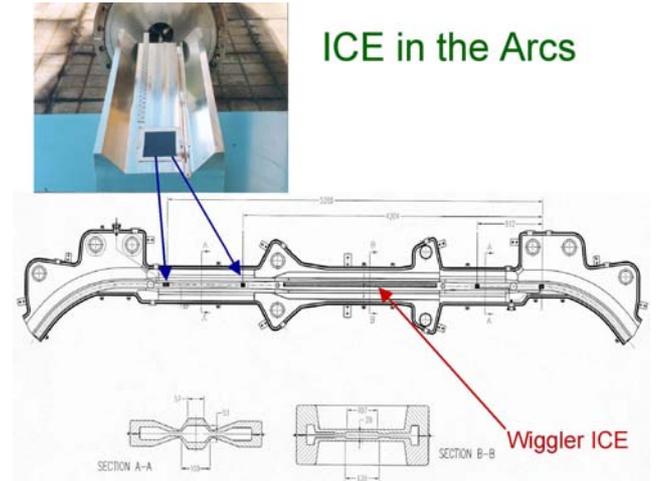

Fig. 3 The arc vacuum chamber with locations of short (bending section) and long (wiggler section) ion clearing electrodes

## 3. BEAM DYNAMICS WITH ICE

Beam measurements have shown that the beam coupling impedance of the two DAΦNE rings were different by approximately a factor of two [6]. The measured impedance of the positron ring is $Z/n$ = 0.54 Ω to be compared with 1.1 Ω of the electron one. This difference produced several harmful consequences affecting the collider performance.

First, at the nominal bunch current of 20 mA the electron bunches were by about 30% longer than positron ones, as shown in Fig. 4. This immediately gives a geometric luminosity reduction due to the well-known hour-glass effect [7]. Moreover, in beam collision schemes with a horizontal crossing angle, as it is the DAΦNE case, synchro-betatron beam-beam resonances become stronger for longer bunches due to a higher

Piwinski angle, thus limiting the maximum achievable beam-beam tune shift parameter.

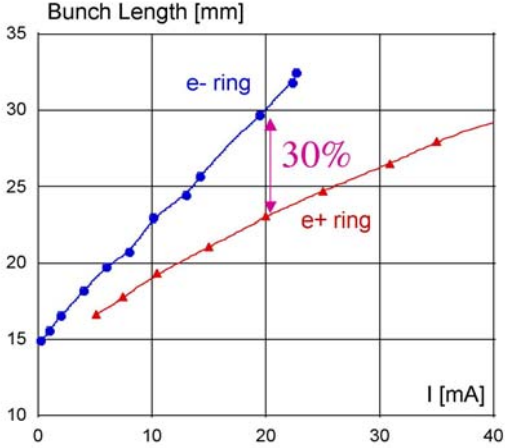

Fig. 4 Bunch length in the electron (blue) and positron (red) rings as a function of bunch current.

Second, the single bunch microwave instability threshold is inversely proportional to the coupling impedance. So, the electron bunches were suffering the instability at lower bunch currents. For DAΦNE the microwave instability leads not only to bunch energy spread widening, but also gives rise to quadrupole (bunch shape) oscillations. This kind of bunch behaviour has been predicted theoretically before DAΦNE commissioning [8] relying on the numerically calculated wake fields [9, 10]. For the positron ring the instability threshold stays above the nominal bunch current [8] while for the electron one it was observed for the currents as low as 9-10 mA in the normal operating conditions. There were several detrimental consequences of that:

- the injection saturation limiting the maximum electron beam current and, respectively, the collider luminosity;
- severe luminosity reduction since in collision the quadrupole oscillations induce additional beam-beam resonances resulting in the beam transverse blow up.

The problem of the quadrupole instability has been solved [11] by tuning the longitudinal feedback system in such way to kick differently heads and tails of bunches. Despite the world record current of 2.4 A has been stored in the DAΦNE electron ring after tuning, further improvements are still possible since in this manner the feedback was working ineffectively due to the necessity of damping at the same time the dipole and the quadrupole coupled bunch oscillations.

The third problem was the vertical beam size blow up (and also the horizontal one, although much weaker) beyond the microwave instability threshold. As shown in Fig. 5, the vertical blow up depends on both the bunch current and the RF voltage $V_{RF}$. At the nominal bunch current and the nominal RF voltage, a 60% blow up is observed even without beam-beam collisions. Besides, a strong correlation has been found between the transverse blow up and the lattice momentum compaction factor $\alpha_c$ [12]. In particular, the lower transverse blow up threshold has not allowed obtaining high luminosity in a collider lattice with a negative momentum compaction factor [13].

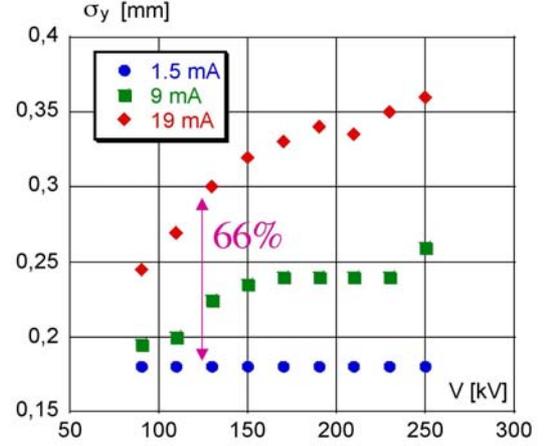

Fig. 5 The vertical beam size measured by the synchrotron light monitor as a function of RF voltage for three different bunch currents

Deeper studies have revealed that the transverse threshold scales accordingly to the Boussard criterion that is used for the longitudinal microwave instability threshold $I_{th}$ estimates [14].

$$I_{th} = \frac{\sqrt{2\pi}\alpha_c (E/e)(\sigma_\varepsilon/E)^2 \sigma_{zo}}{|Z/n|R} \propto \frac{\alpha_c^{3/2}}{|Z/n|\sqrt{V_{RF}}} \quad (3)$$

where $E$ is the beam energy, $\sigma_\varepsilon$ the rms energy spread at zero bunch current, $\sigma_{z0}$ the natural bunch length and $R$ the collider radius.

This has confirmed the dependence of the transverse effect on the longitudinal beam coupling impedance. The vertical size blow up was one of the main factors limiting the DAΦNE luminosity.

## 4. ICE IMPEDANCE

The vacuum chambers of the two DAΦNE rings are similar except for the ICE installed in the electron ring. For that reason and taking into account the relevant impact of the ring impedance on the collider performance we have undertaken more detailed numerical simulations of the ICE impedance [15, 16].

The calculations have shown that the ICE impedance:
- scales linearly with the electrode length $L$ due to the dielectric electrode material;
- scales linearly with the electrode dielectric material thickness $t$;
- scales as a square root of the material dielectric constant $\varepsilon$.

It is worthwhile mentioning that the linear impedance dependence on the electrode length is not a surprise since it is well known that in dielectric channels the wake field builds up proportionally to the channel length. Namely this effect is used in dielectric wake field acceleration experiments (see, for example, [17] and references therein).

It has also been found that the dominant impedance contribution comes from four ICE in the wiggler sections (2.1 m long and very close to the beam). The contribution of all the other ICE is almost negligible since they are much shorter than the wiggler ICE and are located at larger distances from the beam trajectory.

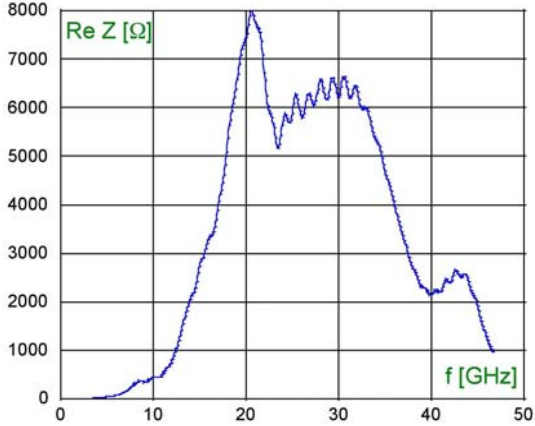

Fig. 6. Real part of the wiggler ICE impedance.

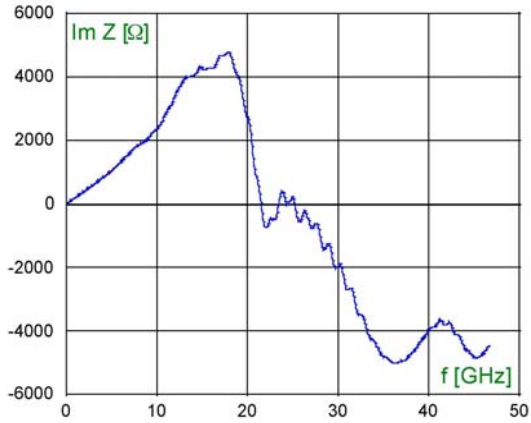

Fig. 7. Imaginary part of the wiggler ICE impedance.

Figure 6 and 7 show the real and the imaginary part of the coupling impedance calculated for the 4 ion clearing electrodes installed in the wiggler sections (4 wigglers per ring). As can be seen, the impedance:
- is broad-band;
- extends till rather high frequencies in the range of tens of GHz;
- is mostly inductive till about 10 GHz.

In turn, Figure 8 shows the respective normalized impedance defined as the beam coupling impedance divided by the revolution harmonic number $n = \omega/\omega o$, where $\omega o$ is the angular revolution frequency. The normalized impedance is almost constant over the large frequency range and accounts for half the electron ring impedance explaining the large impedance difference (about 0.6 Ω) between the electron and the positron DAΦNE rings.

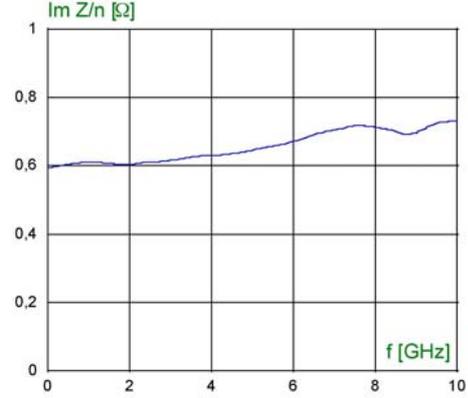

Fig. 8. Normalized impedance of the wiggler ICE.

## 5. BUNCH LENGTHENING IN ELECTRON RING

In order to check the results of the impedance calculations we have performed bunch lengthening simulations and compared their results with available experimental data.

To this purpose we have calculated the wake potential of the 4 wiggler section ICE for a 2.5 mm long Gaussian bunch. The required wake field of the electron ring has been obtained as the sum of the very well known [8] positron ring wake potential and the calculated wake potential of the ion clearing electrodes. In Fig.9 we plot together for comparison the $e^-$ and $e^+$ ring wake potentials.

The above wake potential has been used as a pseudo Green function in the standard tracking code [10] in order to simulate bunch lengthening in the electron ring. Below we compare the tracking results with recent experimental data on the bunch lengthening in the electron ring lattice with negative and positive momentum compaction factors [12, 13].

For the lattice with positive momentum compaction factor the measurements and simulations have been carried out with $\alpha_c = 0.02$ and the RF voltage $V_{RF} = 135$ kV. For each bunch profile acquired by the streak camera and stored in the database we have calculated the rms bunch length. In this way the experimental data can be compared with the rms bunch length given by the simulations. As can be seen in Fig. 10 the agreement is quite satisfactory.

The same procedure has been performed for a lattice with negative momentum compaction factor $\alpha_c = -0.017$ and the RF voltage $V_{RF} = 165$ kV. The full width at half maximum (FWHM) of the longitudinal bunch distribution was measured at different bunch currents and, unfortunately, only a few beam profiles were stored during the measurements. So, in Fig.11 we plot the rms bunch length taken from the simulations and the FWHM/2.35 given by the measurements (we remind here that for a Gaussian bunch FWHM/2.35 is equal to the rms). Only five points (green squares) correspond to the rms bunch length obtained by elaborating the stored bunch profiles. As can be seen in Fig. 11, both simulations and measurements predict the same microwave instability threshold that can be distinguished at the point where the bunch length has its minimum (at about 3 mA per bunch). Figure 12 shows measured and simulated bunch profiles at a bunch current of 7.8 mA.

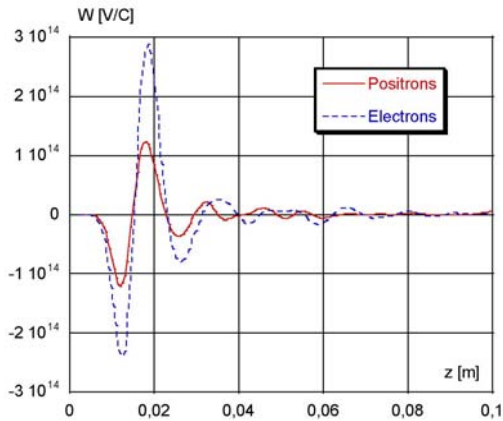

Fig.9. Wake potentials of a 2.5 mm Gaussian bunch for the positron (solid red line) and the electron (dashed blue line) rings.

## 6. ICE REMOVAL AND FIRST RESULTS

Considering the results of the impedance and wake field calculations and the satisfactory agreement between the bunch lengthening numerical tracking and the experimental bunch length measurements it has been decided to remove the ion clearing electrodes from the wiggler sections.

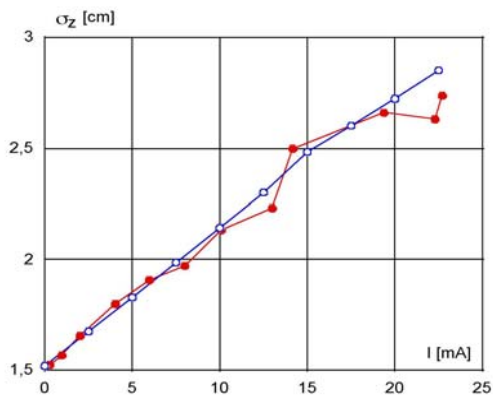

Fig.10. Bunch length in the lattice with positive momentum compaction factor as a function of bunch current: blue dots – numerical tracking; red dots – streak camera measurements.

We did not expect ion trapping worsening after the ICE removal since for the current DAΦNE operating configuration the horizontal emittance is by a factor 3 lower than that in the start-up configuration. Besides, the collider coupling is corrected by a factor 4-5 better with respect to the initial design value. In such conditions only a short gap in the bunch train is required (see Table 1) to expel ions from the electron beam potential.

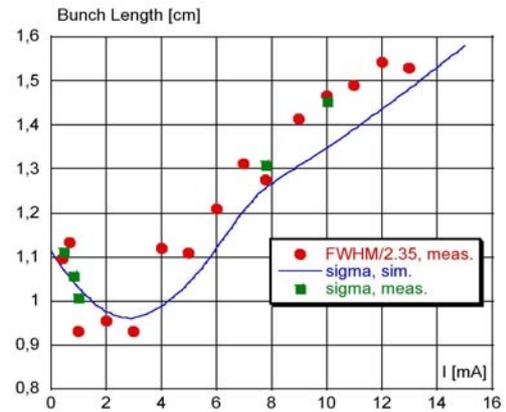

Fig.11. Bunch length in the lattice with negative momentum compaction factor as a function of bunch current: blue line – numerical tracking; red dots – streak camera measurements (FWHM/2.35); green squares – streak camera measurements (elaborated rms).

The ICE removal in-situ was not an easy task because of a limited accessible space overcrowded by other DAΦNE hardware and because of the complexity of the arc vacuum chambers (see Fig.3). The ICE were located in the narrowest part of the chamber which is only 2 cm wide.

In order to extract the ICE from that chamber and to cut the metallic fingers keeping them in place a special dedicated device remotely controlled through an endoscope has been designed and built [18] (see Fig. 13). A kind of a robot moving inside the narrow chamber and equipped with a milling device, a pneumatic piston and a suction cup has been used to both cut the fingers and to extract the electrodes.

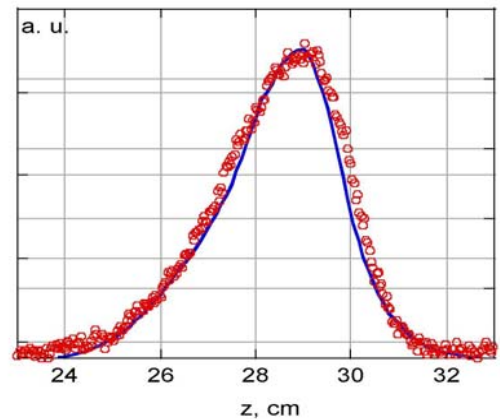

Fig.12. Bunch charge distribution in the lattice with negative momentum compaction factor measured at bunch current of 7.8 mA: red dots – streak camera measurements, blue line – numerical tracking.

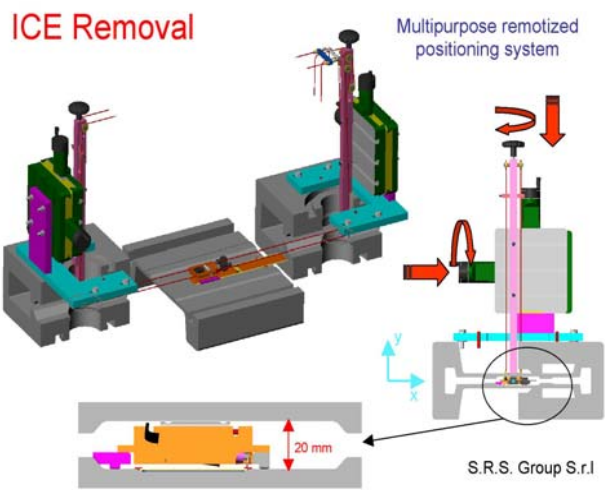

Fig.13. Sketch of the multipurpose remote positioning system used for ICE removal.

The 4 wiggler section ICE have been successfully removed and no vacuum problems have been encountered after DAΦNE re-commissioning. One of the 2 m long ICE is shown in Fig. 14. After the removal observations and measurements, while running the collider for the FINUDA experiment, have confirmed that the electron beam dynamics is now practically similar to that of the positron beam:

- the electron bunches are by about 25-30% shorter for DAΦNE operating conditions;
- the quadrupole instability threshold has been pushed beyond the operating bunch currents;
- no vertical beam size blow up has been observed for the whole range of operating bunch currents and RF voltages.

As can be seen in Fig. 15 in similar working conditions with the same lattice momentum compaction factor and RF voltage (130 kV) bunches are by 25% shorter. Moreover, now we can increase the RF voltage without vertical size blow up. This allows further bunch length reduction by increasing the RF voltage (see the red points at 180 kV, for example).

One of the most important results in terms of collider performance is the geometric luminosity enhancement. Its increase at present is estimated to be around 50%.

## 7. CONCLUSIONS

The experience in running DAΦNE with the "invisible" clearing electrodes has shown that despite the electrode highly resistive layer is transparent for RF frequencies the ICE give a high contribution to the broad-band coupling impedance due to the dielectric material of the electrode plates.

For the DAΦNE electron ring this resulted in several harmful effects: more pronounced bunch lengthening of the electron bunches, lower microwave instability threshold, vertical beam size blow up above the instability threshold and bunch longitudinal quadrupole oscillations.

The detailed wake field and impedance calculations and their experimental verification have revealed that the largest impedance contribution comes from the 4 long electrodes placed in the wiggler vacuum chambers. Their impedance accounts for half the total impedance budget of the electron ring.

The wiggler ICE removal has eliminated the above mentioned harmful effects for the whole range of typical collider operating parameters resulting in a geometric luminosity increase of about 50%.

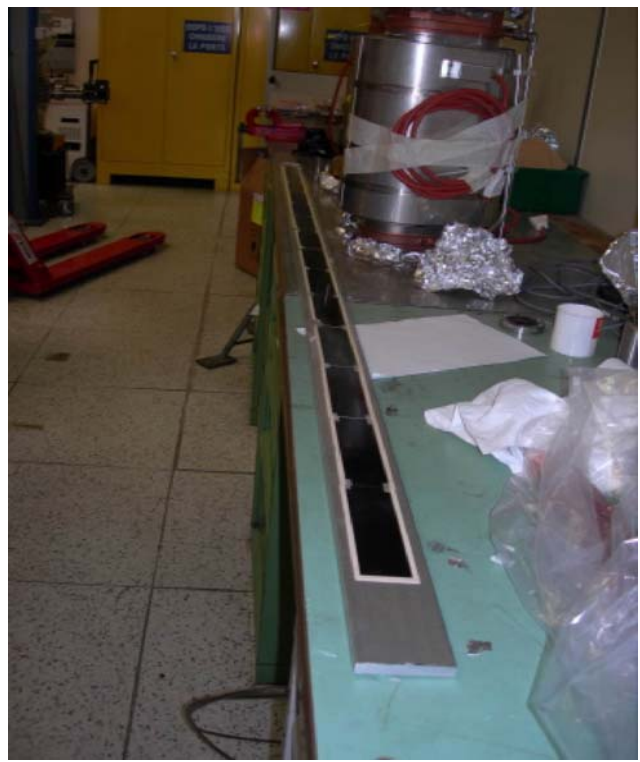

Fig. 14 The ion clearing electrode removed from the wiggler vacuum chamber.

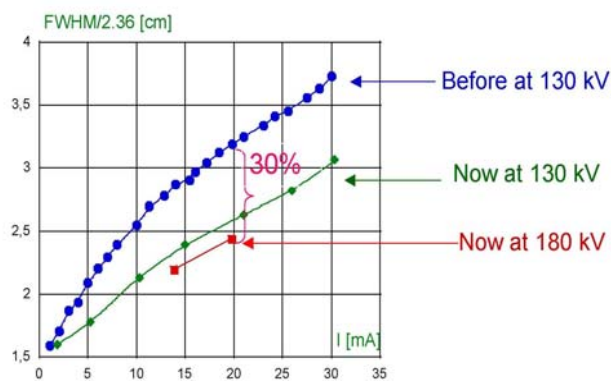

Fig.15 Bunch lengthening before (blue) and after (green) ICE removal. The two red points indicate further bunch length reduction at the higher RF voltage.

## 8. ACKNOWLEDGMENTS

We would like to thank Pantaleo Raimondi for bringing our attention to the ICE impedance problem and Claudio Sanelli for technical supervision of the ICE removal. We are also grateful to David Alesini, Alessandro Gallo and Alessandro Drago for many fruitful discussions and their help in some machine measurements. Eng. Manni is acknowledged for his help in designing the remote positioning system and in removal of the ICE.